\documentclass[11pt]{article}
\usepackage[utf8]{inputenc}
\usepackage[english]{babel}
\usepackage{subcaption} 
\usepackage{graphicx}
\usepackage{amsmath}
\usepackage{upgreek}
\usepackage{amssymb}
\usepackage[unicode]{hyperref}
\usepackage[tableposition=top]{caption}
\usepackage{multirow}
\usepackage{tabularx}
\usepackage{textcomp}
\usepackage{gensymb}
\usepackage{graphics}
\usepackage{color}
\usepackage[dvipsnames]{xcolor}
\usepackage{cleveref}
\usepackage[a4paper, total={6.5in, 8.5in}]{geometry}
\usepackage{tikz}
\usetikzlibrary{backgrounds}

\numberwithin{figure}{section}
\numberwithin{table}{section}
\usepackage{titlesec}
\titleformat{\subsection}{\normalfont\itshape}{\thesubsection}{1em}{}
\usepackage{chngcntr}
\counterwithout{figure}{section}
\counterwithout{table}{section}
\usepackage{soul}
\usepackage{gensymb}

\title{Different configurations of transferred atmospheric pressure plasma jet and their application to polymer treatment}
\date{\today}

\author{Fellype do Nascimento $^{1}$, Antje Quade $^{2}$, Mara A. Canesqui $^{3}$\\and Konstantin G. Kostov $^{1}$\\ \small{\textit{$^{1}$ São Paulo State University -- UNESP, Guaratinguetá, Brazil}}\\ \small{\textit{$^{2}$ Leibniz Institute for Plasma Science and Technology eV, Greifswald, Germany}}\\ \small{\textit{$^{3}$ State University of Campinas -- UNICAMP, Campinas, Brazil}}}

\begin{document}

\maketitle

\begin{abstract}
The employment of atmospheric pressure plasma jets (APPJs) in a large sort of applications is limited by the adversities related to the size of the treated area and the difficulty to reach the target. The use of devices that employ long tubes in their structure has contributed significantly to overcome these challenges. In this work, two different plasma systems employing the jet transfer technique are compared. The main difference between the two devices is how the long plastic tube was assembled. The first one uses a copper wire placed inside a long plastic tube. The other device has a metallic mesh installed in a concentric arrangement between two coaxial plastic tubes. As a result, the two APPJ systems exhibit different properties, with the wire assembly being more powerful, also presenting higher values for the electrical current and rotational temperature when compared to the mesh mounting. X-ray photoelectron spectroscopy (XPS) demonstrates that both configurations were capable of inserting O-containing functional groups on the polypropylene (PP) surface. However, the transferred plasma jet with wire assembly was able to add more functional groups on the PP surface. The results from XPS analysis were corroborated with water contact angle measurements (WCA), being that lower WCA values were obtained when the PP surface presented higher amounts of O-containing groups. Furthermore, the results suggest that the APPJ with wire configuration is more appropriate for material treatments, while the transferred jet with mesh arrangement tends to present lower electrical current values, being more suitable for biological applications.

\end{abstract}

\textit{\textbf{Keywords:} DBD plasma; plasma jets; plasma properties; transferred plasma; plasma treatment; polymer treatment}
\\
\\
\tikzstyle{background rectangle}=[thin,draw=black]
\begin{tikzpicture}[show background rectangle]

\node[align=justify, text width=0.9*\textwidth, inner sep=1em]{
{If you find this preprint useful for your research and want to cite it in your work, please, refer to the published version of that:\\ F. Nascimento, A. Quade, M. A. Canesqui and K. G. Kostov, \textit{Contributions to Plasma Physics}, Vol. 63, e202200055, (2022) -- DOI: \href{https://dx.doi.org/10.1002/ctpp.202200055}{10.1002/ctpp.202200055}}
};

\node[xshift=3ex, yshift=-0.7ex, overlay, fill=white, draw=white, above
right] at (current bounding box.north west) {
\textit{Dear reader,}
};

\end{tikzpicture}
\section{Introduction}
In recent years atmospheric pressure plasma jets (APPJs) produced in open environment have received a lot of attention, not only because of their versatility and ease of use, but also due to the encouraging results achieved in the most diverse applications \cite{brandenburg_dielectric_2017,bekeschus_white_2019, busco_emerging_2020, liu_cold_2020}. Special attention has been given to medical and biological applications of APPJs, whose beneficial effects have been attributed to the reactive oxygen and nitrogen species (RONS) produced by the plasma jets \cite{lu_reactive_2016, cheng_effect_2014, chen_Analysis_2020}. The actions of the RONS are not only limited to biological materials, since they also contribute to the hydrophilization and surface activation processes of a large number of materials treated with APPJs \cite{kim_uniform_2019,nishime_study_2020}.

Most APPJ devices produce plasma jets with small cross-section area, ranging from a several square millimeters to few square centimeters (usually less than 2 cm$^2$). In most cases this is a limitation, since the targets to be treated with APPJs usually have larger dimensions. In other cases, like in dentistry and medicine, it is an advantage because it provides a localized treatment. However, in all situations, it is necessary to provide a way for the plasma to reach the target to be treated. Whether it is bringing the target closer to the device that produces the plasma, or finding a way to bring the plasma to it. For the last purpose, some devices that produce plasma jets at the end tip of long and flexible tubes have been developed \cite{cheng_effect_2014, robert_experimental_2009, polak_innovative_2012, schnabel_multicentre_2012, liu_mri-guided_2014, yousfi_low-temperature_2014, kostov_transfer_2015, liang_plasma_2015, borges_amplitude-modulated_2018, geng_flexible_2018, omran_atmospheric_2018, tang_new_2018, Winter_development_2018, kurosawa_endoscopic_2019, schweigert_interaction_2019, bastin_optical_2020, chen_transferred_2021}. Being that some of them use the plasma jet transfer technique, which is basically a secondary plasma jet formed away from a primary discharge that is generated inside a reactor or from a primary plasma jet \cite{cheng_effect_2014, kostov_transfer_2015, borges_amplitude-modulated_2018, bastin_optical_2020, xiong_atmospheric-pressure_2013, xia_transfer_2016}. One of the advantages of use the jet transfer technique in APPJ devices with long tubes is that the plasma jet is produced far from the high voltage source which improves the electrical safety in its operation and handling. Furthermore, the use of the jet transfer technique facilitates the employment of an APPJ to treat large target materials by scanning their surfaces.

One of the problems related to the generation of plasma jets using long polymeric tubes is the possibility of plasma production in the tube length instead of obtaining a plasma jet only at the end tip of the tube. This is a problem because the plasma tends to interact with the tube material causing a degradation. Some studies found that this can be prevented by using a second tube surrounding the first, with a space to flow a different gas between the two tubes \cite{Winter_development_2018, Winter_enhanced_2018}. A study by Onyshchenko \textit{et al} aimed to investigate penetration depth of plasma inside flexible tubes suggests that this problem can be minimized by employing long tubes with an appropriated internal diameter \cite{Onyshchenko_atmospheric_2015}. In that work it was observed that the plasma penetration depth in tubes depends on their internal diameters, being that there is a certain tube diameter that maximizes the penetration depth.

The devices that employ long and flexible tubes to produce plasma jets for materials treatment have shown results that are as good as those that do not use long tubes. 
However, the most interesting results are related to the possibility of performing treatments in cavities, specially in combination with endoscopes \cite{Winter_development_2018, kurosawa_endoscopic_2019, Winter_enhanced_2018}. Nonetheless, before to apply a new configuration or technique to produce APPJ on biological materials or on \textit{in vivo} targets, it is convenient, and also a common practice, to begin studies with inorganic materials like polymers. 

In this work we present a modified design concept using the jet transfer to generate APPJs at the ends of long and flexible tubes. It consists of a metallic mesh placed between two long plastic tubes in a concentric arrangement. The main objective of using this design is to reduce the electrical current that flows through the plasma jet thus making it more suitable for \textit{in vivo} applications. This configuration was compared with another that employs a metal wire inside a long tube. Both devices were characterized with electrical and optical diagnostics and comparisons of the plasma parameters obtained on each case are provided. A study comparing the use of the two different configurations for polymer treatment was also performed. The results revealed differences in the surface treatments when each configuration is used.

\section{Experimental setup}
\begin{figure}[htb]
\centering
\begin{subfigure}{0.44\textwidth}
\centering
\includegraphics[width=\textwidth]{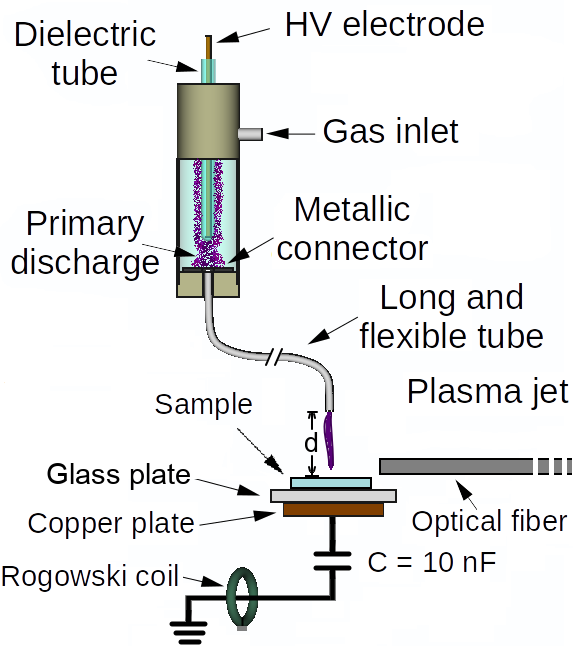}
\caption{Setup overview}
\end{subfigure}\quad
\begin{subfigure}{0.53\textwidth}
\centering
\includegraphics[width=\textwidth]{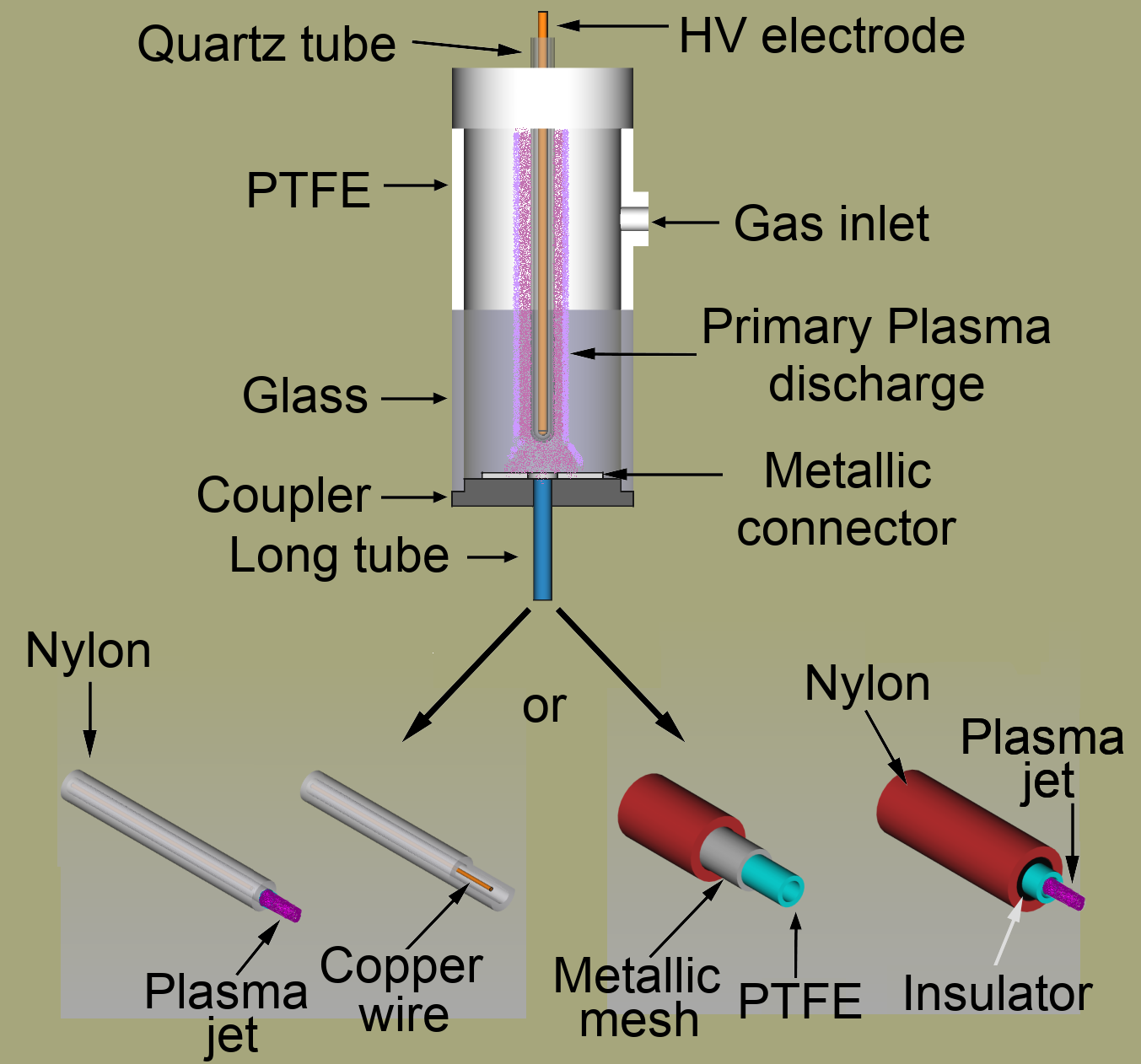}
\caption{Schemes of the two long tubes}
\end{subfigure}
\caption{(a) Overview of the experimental setup. (b) Details of the two long tube configurations.\label{expset}}
\end{figure}

Figure~\ref{expset}(a) shows the experimental setup used in this work, with a detailed view of the two long tube configurations illustrated in Fig.~\ref{expset}(b). The plasma jet device presented in Fig.~\ref{expset} is based on a DBD type reactor, composed by a metal pin electrode encapsulated in a closed-end quartz tube, which in turn is placed inside a dielectric chamber. Completing the device, there is a 1-meter long and flexible plastic tube connected to the DBD reactor.
In this work, the long tube employed to produce APPJ was assembled in two different ways. In the first one a 0.5 mm thin copper wire is put inside a plastic tube, whose material is Nylon 6, with outer diameter (OD) equals to 4 mm and inner diameter (ID) equals to 2 mm. In the second configuration the long tube is composed by a metallic mesh, with 90$\%$ closed area extended over a polytetrafluoroethylene (PTFE) tube with OD = 3 mm and ID = 2 mm, being that both of which are placed inside a Nylon 6 tube with OD = 6 mm and ID = 4 mm.

Both the wire and the mesh are fixed to a metallic connector that goes inside the reactor and acts as a floating electrode (see Fig.~\ref{expset}). Besides that, each set formed by the long tube and metallic connector is attached to its own exchangeable coupler, which is made of polyoxymethylene (POM). Furthermore, both copper wire and metallic mesh ends 2 mm before the plasma outlet, and in the last configuration there is an electrical insulation at the end of the mesh in order to avoid discharges coming directly from it.

In order to produce plasma the working gas, argon (Ar) for instance, is fed into the chamber and flows through the long and flexible plastic tube connected to the reactor. At the same time a high voltage (HV) signal is applied to the HV electrode and a primary discharge is formed inside the reactor chamber. When the primary discharge is on, it polarizes the metallic connector, which is linked to the copper wire or metallic mesh, and a small plasma jet is ignited at the distal end of the plastic tube. The remote plasma jet produced in each case was directed at normal incidence towards a target, placed at a distance \textbf{d} = 6 mm from the plasma outlet. We performed the electrical and optical characterizations of the plasma jets using two different targets. One of them was a dielectric material (the glass plate in {Fig.~\ref{expset}}(a)) lying over a grounded copper plate, and the other was the copper plate itself.

\begin{figure}[htb]
\centering
\includegraphics[width=10 cm]{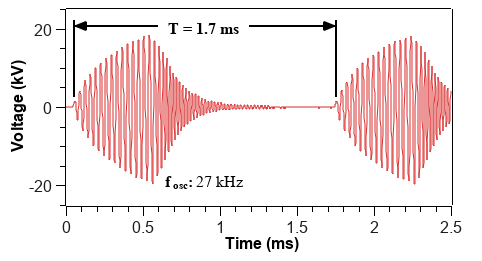}
\caption{Example of a typical HV waveform applied to the pin electrode without producing plasma. \label{HVwform}}
\end{figure}

The power source employed to produce the plasma jets was a commercial AC generator from GBS Elektronik GmbH (model Minipuls4). We choose to apply a HV waveform that resembles a sinusoidal ``burst'' instead of a pure sinusoidal signal because the former produces lower ohmic heating on the DBD reactor and on other parts of the device. The oscillation frequency ($f_{osc}$) in the burst HV was 27 kHz. {This $f_{osc}$ value was chosen as the one} that produces the most stable plasma jets in both wire and mesh configurations. The HV burst is followed by a voltage off period, which repeats at a repetition period ($\tau$) of 1.7 ms. An example of the typical HV waveform applied to the powered electrode is depicted in Fig.~\ref{HVwform}.

Simultaneous measurements of the voltage applied on the powered electrode and the voltage across a serial capacitor (C = 10 nF) were carried out in order to obtain the average discharge power ($P_{dis}$). The calculation of $P_{dis}$ values takes into account all voltage oscillations in each burst that produce a charge variation in C. The applied voltage was measured using a 1000:1 voltage probe (Tektronix, model P6015A) and the waveforms were recorded using a 200 MHz oscilloscope (Tektronix, model 2024B). Then, the $P_{dis}$ value is calculated by summing the area of the $q-V$ Lissajous figures formed between the voltage ($V(t)$) and charge ($q(t)$) signals, divided by the burst period $\tau$, that is \cite{holub_measurement_2012,ashpis_progress_2017, pipa_equivalent_2019}:

\begin{equation}
{P_{dis} = \dfrac{1}{\tau} \oint q(V) dV}\label{eqpower}
\end{equation}

By applying the Green's theorem to (\ref{eqpower}), $P_{dis}$ can be calculated using:

\begin{equation}
{P_{dis} = \dfrac{1}{2 \tau} \int_{t_1}^{t_2} \left[ V(t) q'(t) - V'(t) q(t) \right] dt}\label{eqpgreen}
\end{equation}

\noindent where $\{V'(t),q'(t)\} = d\{V(t),q(t)\}/dt$. An advantage of using Eq.(\ref{eqpgreen}) to calculate the area of the Lissajous figure is that it can be done without the need to plot the $q-V$ curve.

We also measured the waveform of the current that passes through the system using a Rogowski coil from Pearson${}^{\rm{TM}}$ (model 4100), which was used to obtain the effective discharge current ($i_{RMS}$).

To observe the light emission from multiple excited species and evaluate the production of the OH and NO species in the plasma, a broad-band optical emission spectroscopy (OES) in the wavelength range from 200 nm to 750 nm was performed using a multi-channel spectrometer from Avantes (model AvaSpec-ULS2048X64T), with spectral resolution (FWHM) equal to 0.76 nm. The light emitted by the plasma jet was collected using an optical fiber placed close to the surface target. The distance between the center of the plasma column and the fiber optic light input was 5 mm.

Spectroscopic measurements were also used to obtain rotational and vibrational temperatures ($T_{rot}$ and $T_{vib}$, respectively) of $\rm{N_2}$ molecules. In order to obtain $T_{rot}$ and $T_{vib}$ values, we used spectral emissions from the $\rm{N_2}$ second positive system, $C {}^{3} \Pi_u, \nu' \rightarrow B {}^{3} \Pi_g , \nu''$, with $\Delta \nu$ = $\nu' - \nu''$ = -2, in the wavelength range from 362 to 382 nm \cite{moon_comparative_2003,bruggeman_gas_2014,zhang_determination_2015,ono_optical_2016}. Spectra simulations were carried out using the massiveOES software \cite{vorac_batch_2017,vorac_state-by-state_2017}. Thus, comparisons between measured and simulated spectra were performed and the temperature values that generate a simulated curve that best fit to the experimental spectrum are chosen as the $T_{rot}$ and $T_{vib}$ values of the plasma jet. It is known that spectroscopic measurements performed with low-resolution spectrometers, like the Avantes one, are not sufficient to fully resolve the rotational levels of the $\rm{N_2}$ molecules, which is a requirement to obtain accurate values for the $T_{rot}$ parameter. However, there is a direct relationship between the shape and broadening of the $\rm{N_2}$ vibrational bands and the variation of the $T_{rot}$ values, being that the higher the $T_{rot}$, the larger the broadening and also the higher the intensity of the rotational lines in the vibrational bands. Both effects cause a change in the shape of the vibrational bands in that part that degrades to violet, causing it to become higher and wider, allowing the estimation of the $T_{rot}$ values by using low-resolution spectrometers. Thus, even not very accurate, the $T_{rot}$ values measured with the Avantes spectrometer can be good enough to show the trend of that parameter in the experiments performed in this work.

In this work we chose to perform the spectroscopic measurements in close proximity to the target surface because this is the portion of plasma that enters in contact with the material under treatment. So, this seems to be the most appropriate measurement for possible observations of the effects of the target's material on the plasma jets. However, when performing such measurements close to the surface of the target and using the data to measure the rotational temperatures, the $T_{rot}$ values obtained may be different from the $T_{gas}$ ones \cite{bruggeman_gas_2014}.

The APPJs produced with both wire and mesh configurations were also applied on polypropylene (PP) samples as a way to evaluate the effects of the plasma properties on the treatment of materials. Those effects were analyzed mainly by measurements of water contact angle (WCA), X-ray photoelectron spectroscopy (XPS) and morphology studies. The XPS measurements were performed with an equipment from Kratos (model Axis Supra), whose energy resolution is better than 1 eV on polymers and the local resolution is better than 15 $\upmu$m. WCA measurements were carried out with a goniometer from Ramé-Hart (model 300), using the sessile drop method. Surface morphology analysis were performed using an optical profiler from Nanovea (model PS50).

The PP samples used in this work are commercially available materials and present high density values. All the samples used have 1-mm thickness and approximately the same dimensions (35 mm x 25 mm).

A simple test for evaluating the surface temperature of a thin copper target (20 mm x 45 mm, 0.5 mm-thick) upon plasma jet bombardment for 10 min was performed. The jet was operated with Ar gas at flow rates of 1.0 and 2.0 l/min at a distance of 6 mm to the target. The equipment used for those temperature measurements was an infrared thermometer from Minipa (model MT-395).

\section{Results and discussion}

\subsection{Discharge electrical parameters for the different configurations}\label{section:ElectricalPar}

\begin{figure}[htb]
\centering
\begin{subfigure}{0.49\textwidth}
\centering
\includegraphics[width=\textwidth]{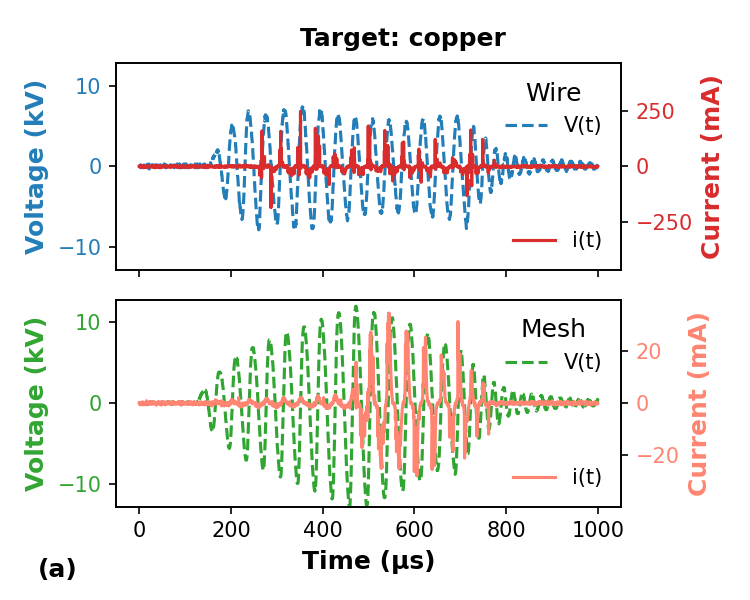}
\end{subfigure}
\begin{subfigure}{0.49\textwidth}
\centering
\includegraphics[width=\textwidth]{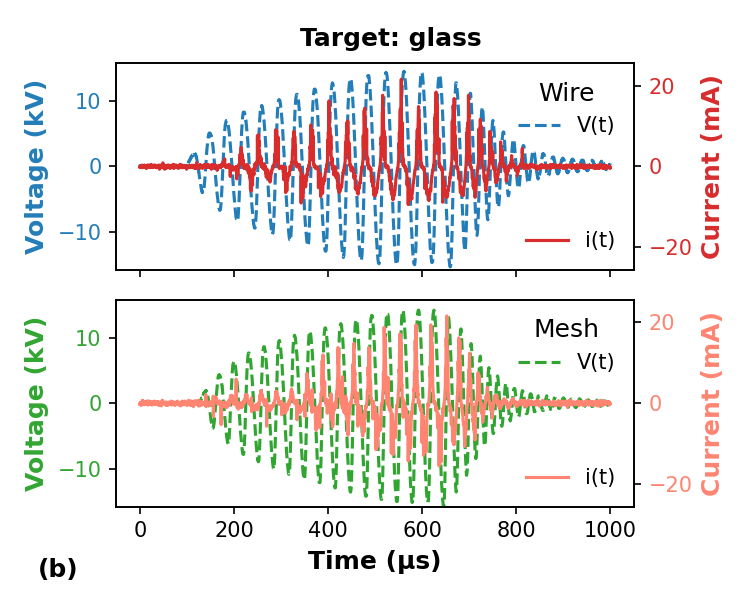}
\end{subfigure}
\caption{Typical waveforms of applied voltage (V(t)) and discharge current (i(t)) using (a) copper and (b) glass as targets, for both wire and mesh configurations. The gas flow rate was 2.0 l/min in all cases. \label{WVforms}}
\end{figure}

In Fig.~\ref{WVforms} are shown the typical voltage and current waveforms measured using (a) a metallic (copper) target and (b) a dielectric (glass) one (b) for both wire and mesh configurations. In the condition where the APPJ impinges on the copper target it can be observed that when we switch from wire to mesh configurations there is an increase in the peak value of the applied voltage. However, what we have in fact is that the use of the wire configuration to produce the APPJ causes a voltage drop on the applied voltage, since the power source is configured to deliver always the same energy to the APPJ device, independently of the configuration in use. In other words, the peak voltage value is always the same ($\sim$15 kV-peak) when there is no plasma discharge in both configurations, and it looks like the waveform shown in Fig.~\ref{HVwform}. We kept the nominal voltage constant while the actual applied voltage may vary depending on the device configuration and target conductivity. In this study we intend to evaluate the effect of the configuration (wire or mesh) on the electrical parameters. Furthermore, that voltage drop is strongly linked to the higher electrical current values obtained in the wire configuration. Nevertheless, the voltage drop obtained with the wire configuration is much lower when we use the glass plate as the target, which confirms the latest statement. In addition, with the glass target there is no significant differences in the voltage drop when we switch from the wire configuration to the mesh one.

Figure~\ref{WVforms} also shows that there is a clear ignition delay and a shorter discharge duration when the mesh is used in comparison with the wire configuration. Such discharge ignition delay is less evident when the glass target is used.
Regarding the amplitude of current signals shown in Fig.\ref{WVforms}, it can be seen that when the plasma jet impinges on the copper target the peak values are much higher (by one order of magnitude) in the wire configuration than in the mesh one. However, the amplitude of the current signals are of the same magnitude order when the plasma jet impinges on the glass target.

The discharge power ($P_{dis}$) and the electrical current, presented in this work as the root mean square value of the current waveform ($i_{RMS}$), are the most important electrical parameters in APPJs. Usually, it is desirable to have APPJs with adequate power values and low electrical current. Being that the last condition is highly desired for \textit{in vivo} applications. However, plasma jets with low $P_{dis}$ values are also useful when sensitive materials are under APPJ treatment.

\begin{figure}[htb]
\centering
\begin{subfigure}{0.49\textwidth}
\centering
\includegraphics[width=\textwidth]{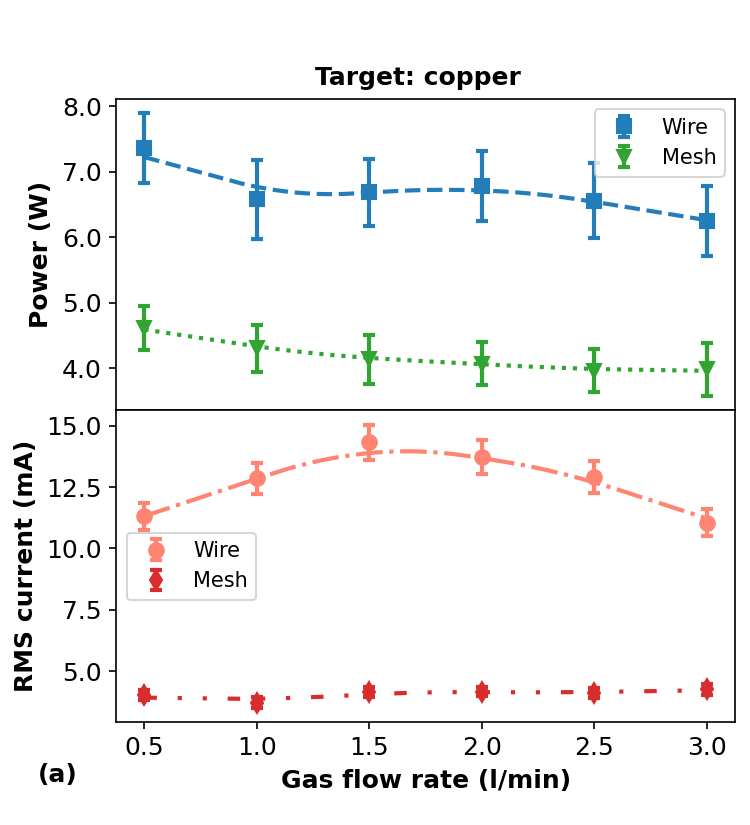}
\end{subfigure}
\begin{subfigure}{0.49\textwidth}
\centering
\includegraphics[width=\textwidth]{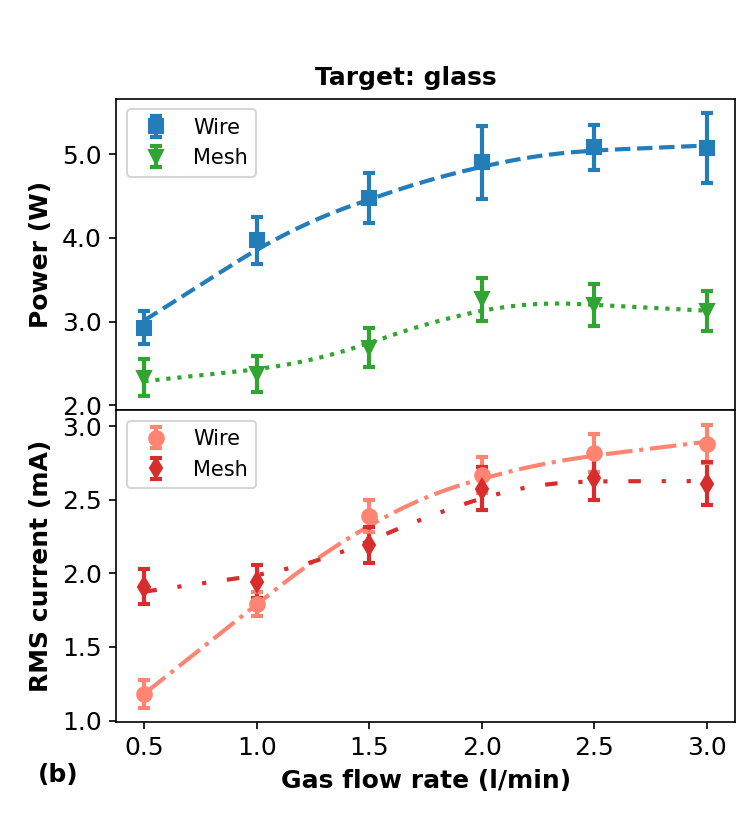}
\end{subfigure}
\caption{Discharge power and effective current as a function of the gas flow rate using (a) copper and (b) glass as target. \label{piVSfRate}}
\end{figure}

Figure~\ref{piVSfRate} shows the curves of $P_{dis}$ and $i_{RMS}$ as a function of the gas flow rate ($Q$) for a fixed distance (d = 6 mm) between the plasma outlet and the target. The results obtained for the metallic and the dielectric targets are shown in Figs.~\ref{piVSfRate}(a) and ~\ref{piVSfRate}(b), respectively. For both targets the curves of $P_{dis}$ and $i_{RMS}$ vs $Q$ were measured for the two 1-meter long tube configurations studied in this work.
The first noticeable difference when we compare the wire and the mesh configurations is that the $P_{dis}$ curves present higher values in the wire configuration. Considering the copper target, the two $P_{dis}$ curves present similar trends, decreasing as the gas flow rate increases, being that the $P_{dis}$ values obtained with the wire are, in general, $\sim$60$\%$ higher than those obtained with the mesh. On the other hand, the $i_{RMS}$ curves present different behaviors. When the wire is used, the $i_{RMS}$ curve exhibits a non-monotonic behavior with a peak value at $Q \approx$ 1.5 l/min, while the $i_{RMS}$ values are almost constant when using the mesh configuration. Also, the average $i_{RMS}$ values are nearly three times higher when the wire is employed.

The explanation for the lower values of $P_{dis}$ and $i_{RMS}$ obtained with the mesh configuration when compared to the wire one is quite simple: the PTFE tube inside the mesh electrode acts as an additional dielectric barrier. So, this can be used as a way to improve the electrical safety of the equipment, making it more suitable for handling and \textit{in vivo} applications. Of course the shorter discharge duration achieved when the mesh is employed contributes significantly to the reduction of both $P_{dis}$ and $i_{RMS}$ values.

Analyzing the results obtained with glass as the target, it can be seen that in general both $P_{dis}$ and $i_{RMS}$ curves present similar trends as a function of $Q$ in the wire or mesh configurations. The average $P_{dis}$ values are $\sim$56$\%$ higher in the wire configuration when compared to the mesh one, which is almost the same increment obtained with copper as the target. The $i_{RMS}$ curves differ significantly only for $Q$ values lower than 1.0 l/min, with a small decrement when the wire is used and a small increment in the mesh case, but, in general, the $i_{RMS}$ values are almost the same.

\subsection{Spectroscopic emissions and thermal parameters}\label{section:Spectroscopy}
\begin{figure}[htb]
\centering
\begin{subfigure}{0.49\textwidth}
\centering
\includegraphics[width=\textwidth]{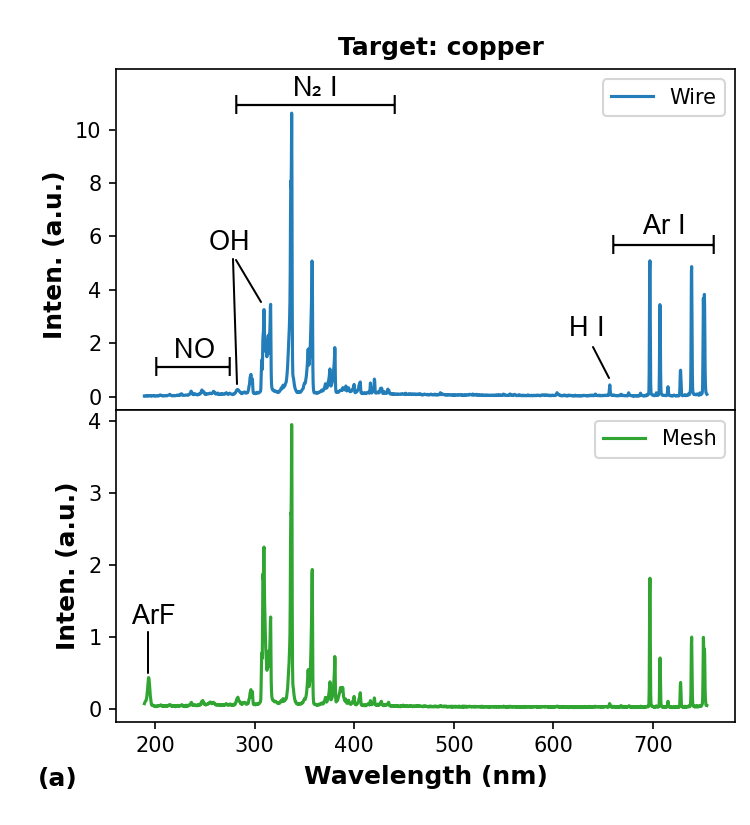}
\end{subfigure}
\begin{subfigure}{0.49\textwidth}
\centering
\includegraphics[width=\textwidth]{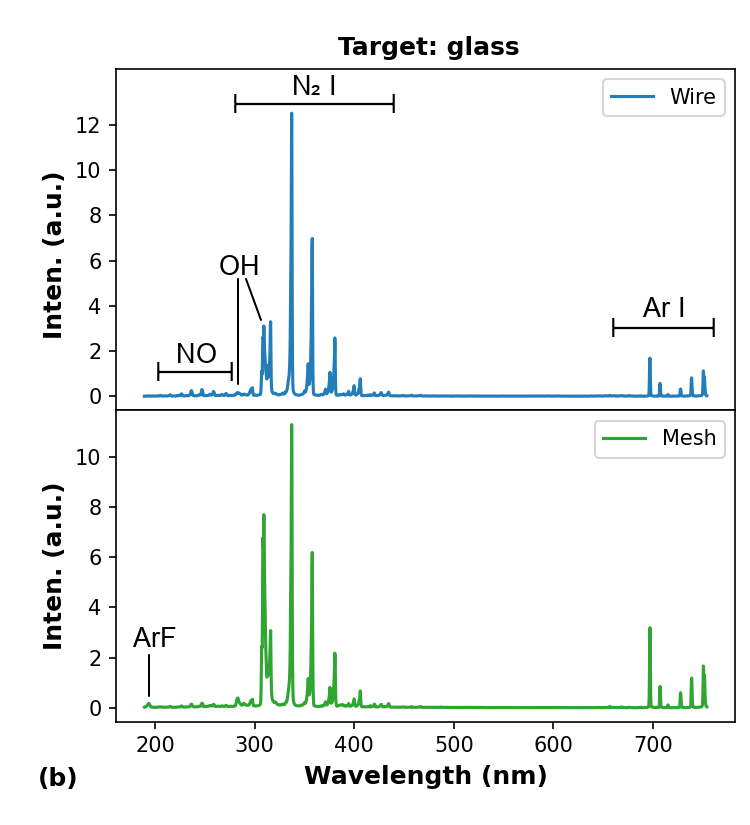}
\end{subfigure}
\caption{Overview of the emission spectra of the plasma jets obtained for the two different configurations with (a) copper and (b) glass as targets. The gas flow rate was 2.0 l/min in all cases. \label{spectraov}}
\end{figure}

Figure~\ref{spectraov} shows an overview of the emission spectra obtained with Ar as working gas for the plasma jets produced in the wire and mesh configurations and using the copper target or the glass one. All spectra shown in Fig.~\ref{spectraov} exhibit emissions from NO, in the 200 nm to 270 nm wavelength range, OH at 288 nm, 296 nm and 308 nm, with the last two bands jeopardized by $\rm{N_2}$ emissions, and from $\rm{N_2}$ molecules ($\rm{N_2}$ I) in the 298-450 nm range. The hydrogen alpha emission line at 656.28 nm is detected only in the spectra obtained using the copper target. Since the glass target is in fact an additional dielectric barrier this leads to a reduction of both the electron energy and density and, consequently, the probability of H atoms excitation by electron impact becomes lower. Argon line emissions were also observed in all spectra. Regarding the configurations used, there are no significant differences in the emitting species when the remote plasma jet is produced using wire or mesh. The only noticeable difference is the appearance of an ArF molecular emission close to 193 nm when the mesh configuration is employed. That ArF emission was observed previously in an APPJ spectrum by Polak \textit{et al} \cite{polak_innovative_2012}, and its appearance is probably due to the interaction between the Ar plasma jet with the wall of the PTFE tube over which the metal mesh is wrapped.

\begin{figure}[tb]
\centering
\begin{subfigure}{0.49\textwidth}
\centering
\includegraphics[width=\textwidth]{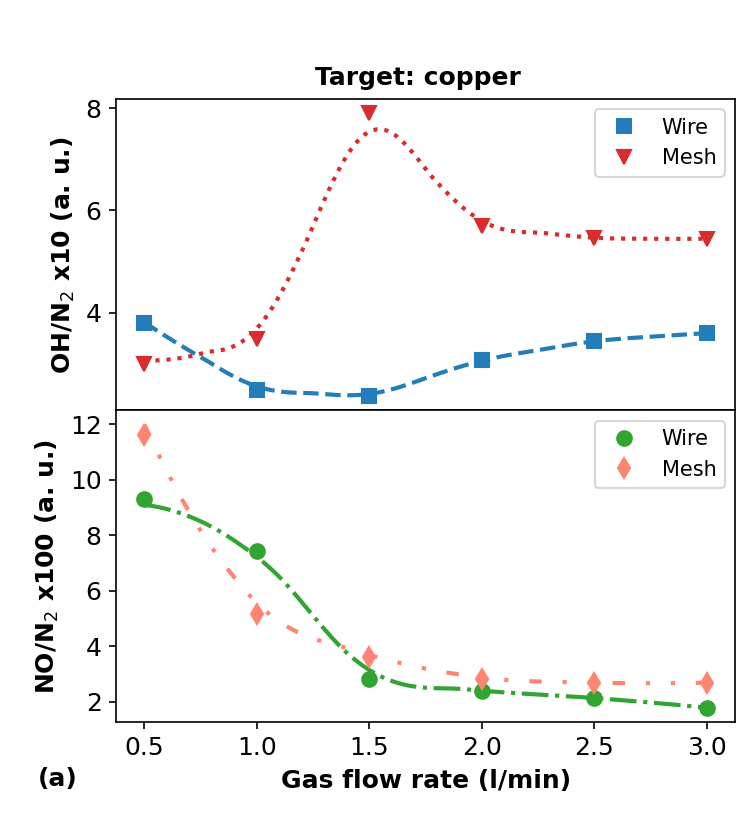}
\end{subfigure}
\begin{subfigure}{0.49\textwidth}
\centering
\includegraphics[width=\textwidth]{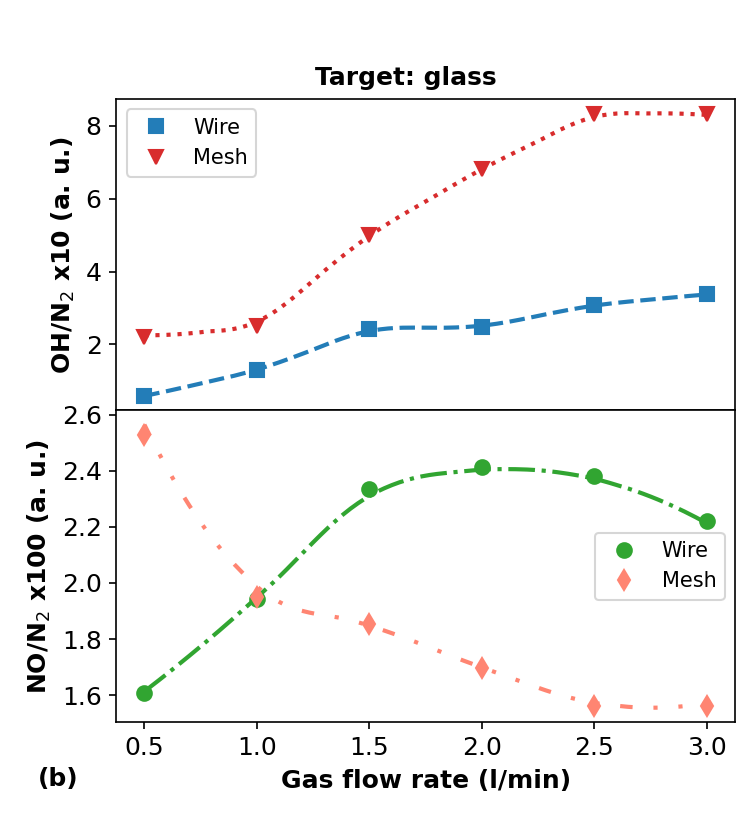}
\end{subfigure}
\caption{OH/$\rm{N_2}$ and NO/$\rm{N_2}$ ratios as a function of the gas flow rate for (a) copper and (b) glass targets. \label{iratiosVSQ}}
\end{figure}

An important observation to be made about the spectra shown in Fig.~\ref{spectraov} is related to the relative intensity of the different emitting species. Although the overall emission intensity with the mesh is low, the relative intensities of OH and NO emissions with respect to N$_{2}$ change considerably. The curves for the ratios between the intensity emissions of OH and NO with respect to $\rm{N_2}$ (OH/$\rm{N_2}$ and NO/$\rm{N_2}$, respectively) as a function of $Q$ are shown in Fig.~\ref{iratiosVSQ}. The wavelengths of the emitting species used to calculate the ratios shown in Fig.~\ref{iratiosVSQ} were 247 nm for NO, 308 nm for OH and 337 nm for $\rm{N_2}$. The production of both OH and NO species in Ar plasma jets involves collisions of $\rm{N_2}$ in metastable states, which in turn are generated by electron-impact and also by collisions with Ar atoms in excited and in metastable states \cite{Gaens_reaction_2014,Ghimire_influence_2021}. In addition, the emission intensity of the plasma species is proportional to their densities. Thus, the intensity ratios OH/$\rm{N_2}$ and NO/$\rm{N_2}$ provide a rough estimate about the efficacy in the production of OH and NO in Ar plasma jets.

From Fig.~\ref{iratiosVSQ}(a), it can be seen that for the conducting target the OH/$\rm{N_2}$ curves behave differently when the wire or the mesh is used. Regarding the ratio values observed in Fig.~\ref{iratiosVSQ}(a), we see that the OH/$\rm{N_2}$ ratio present higher values using the mesh configuration instead the wire one. Furthermore, from {Fig.~\ref{iratiosVSQ}(a)}, we observe that when the APPJ impinges on the copper target the mesh configuration tends to provide an optimal OH/$\rm{N_2}$ ratio when the gas flow rate is close to 1.5 l/min and that the NO production in relation to $\rm{N_2}$ tend to decrease as $Q$ increases.

When the glass target is employed the behavior of OH/$\rm{N_2}$ curves shown in Fig.~\ref{iratiosVSQ}(b) present the same trend when wire or mesh configuration are used. A point that is noticeable here is that the OH/$\rm{N_2}$ ratio tends to be higher when the mesh configuration is employed. Regarding the NO/$\rm{N_2}$ ratio, the behavior of the curves shown in Fig.~\ref{iratiosVSQ}(b) are quite different when the configuration is changed. Being that in the case of wire configuration the NO/$\rm{N_2}$ present a nearly parabolic shape with a peak value at $Q \approx 2.0$ l/min. On the other hand, the NO/$\rm{N_2}$ curve obtained with the mesh mounting decrease monotonically as $Q$ is increased.

By comparing the OH/$\rm{N_2}$ intensity ratios obtained with both wire and mesh configurations it can be noticed that the mesh one tends to be more efficient in the OH production, regardless of target material. On the other hand, the NO/$\rm{N_2}$ intensity ratios reveal that when the target material is copper the efficacy in generating NO is almost the same. However, for the glass target the wire mounting is clearly more efficient than the mesh one for higher gas flow rates, while the opposite occurs for very low flow rates.

It is important to mention that the uncertainty in the intensity measurements are of the order of 1$\%$, and the repeatability of measurements changes no more than 3$\%$ under the same environmental conditions. Then, the uncertainties in the intensity ratio calculation are unlikely to cause a significant change in the behavior of the curves shown in {Fig.~\ref{iratiosVSQ}}.
\\

Regarding the thermal parameters of the APPJs produced using the configurations under study, we measured only the rotational and vibrational temperatures ($T_{rot}$ and $T_{vib}$, respectively). These two thermal parameters can be considered the main ones for APPJs, since $T_{vib}$ is related to the rate of chemical reactions, and $T_{rot}$ has a close relationship with the gas temperature ($T_{gas}$), with $T_{rot} \approx T_{gas}$ in most cases \cite{moon_comparative_2003,bruggeman_gas_2014,lambert_vibrationvibration_1967, smith_preference_2004}.

\begin{figure}[htb]
\centering
\begin{subfigure}{0.49\textwidth}
\centering
\includegraphics[width=\textwidth]{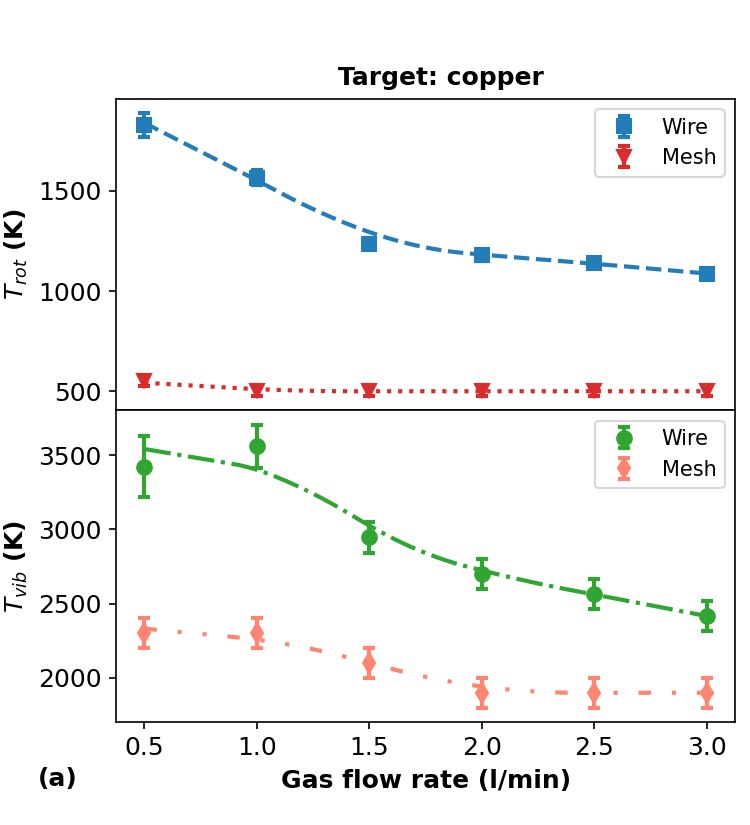}
\end{subfigure}
\begin{subfigure}{0.49\textwidth}
\centering
\includegraphics[width=\textwidth]{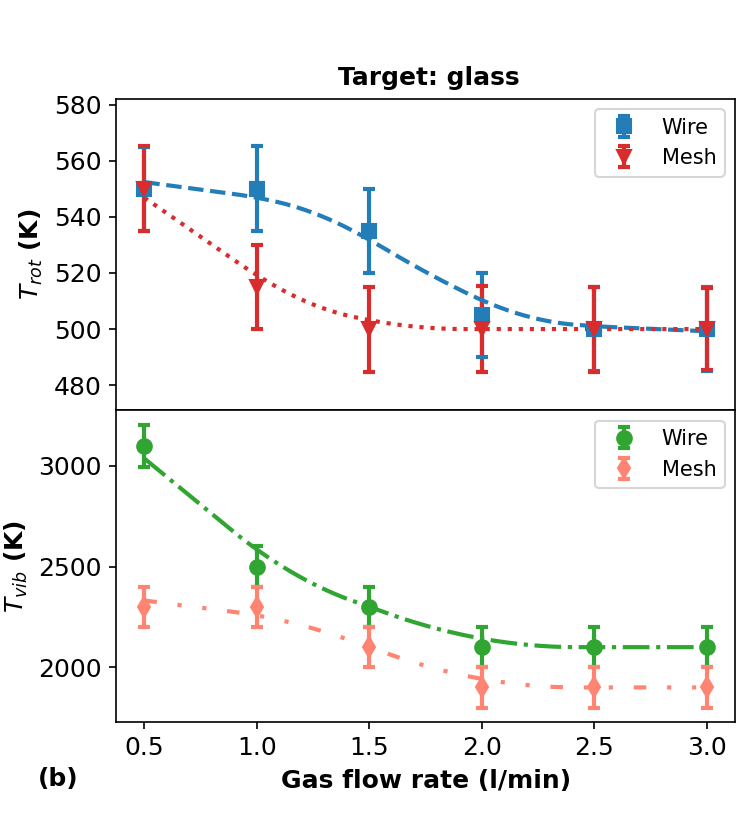}
\end{subfigure}
\caption{Rotational and vibrational temperatures as a function of the gas flow rate for (a) copper and (b) glass targets. \label{tempsVSfRate}}
\end{figure}

Figure~\ref{tempsVSfRate} shows the curves of $T_{rot}$ and $T_{vib}$ as a function of $Q$, obtained using the two different targets, for the wire and mesh configurations.
From Fig.~\ref{tempsVSfRate}(a) it can be seen that when the copper plate is used as the target both $T_{rot}$ and $T_{vib}$ values decrease significantly when we switch from the wire to the mesh configuration. Regarding the $T_{rot}$ values, the reduction observed is more than three times for $Q \leq$ 1.0 l/min and more than the double for higher $Q$ values. The proportion in the reduction of $T_{vib}$ was of the order of 50$\%$ for $Q \leq$ 1.0 l/min with the difference becoming smaller for other $Q$ values.

On the other hand, when the APPJ impinges on the dielectric target both $T_{rot}$ and $T_{vib}$ values do not differ significantly between the wire and mesh configurations, presenting also the same trend, that is, decrease in temperature values as $Q$ is increased. Nevertheless, the $T_{vib}$ values tend to be slightly higher when the wire mounting is employed, and even more for $Q <$ 1.0 l/min.

It is important to stress at this point that the statement $T_{rot} \approx T_{gas}$ is not always valid when the $T_{rot}$ measurements are performed using spectral emissions from the $\rm{N_2}$ ($C {}^{3} \Pi_u, \nu' \rightarrow B {}^{3} \Pi_g , \nu''$) system. This finding was discussed extensively in a review paper by Bruggeman \textit{et al} {\cite{bruggeman_gas_2014}}. In our work, in particular, the spectroscopy measurements were performed close to the surface of the target and the working gas was argon, which are two conditions in which we have $T_{rot} \neq T_{gas}$, according to {\cite{bruggeman_gas_2014}}, because when Ar is the working gas the population and excitation of rotational levels of the $\rm{N_2} (C)$ vibrational band are not dominated by electron impact collisions, but by collision with Ar atoms and $\rm{N_2}$ molecules in metastable states, which can produce $\rm{N_2} (C)$ excited states with specific rotational distributions. In addition, when the excitation of the $\rm{N_2} (C)$ state occurs by collisions with Ar metastables its rotational levels tend to be largely excited. In this case the APPJ may present $T_{rot}$ values higher than $T_{gas}$.
Regarding the measurements performed close to a surface, collisions of the gas components with a surface also influence the rotational distribution, most times reducing the rotational excitation, which can lead to a reduction in $T_{rot}$ values and to an underestimation of $T_{gas}$ if both are assumed to have the same value.
However, we prefer to perform the spectroscopic measurements close to the target surface because this can provide a better insight about the plasma-material interaction. From the $T_{rot}$ curves presented in {Fig.~\ref{tempsVSfRate}}, it can be seen clearly that when the wire configuration is used the employment of the conducting target leads much higher to $T_{rot}$ values than in the case of the dielectric substrate. This probably happens because the plasma jet tends to enter in a spark regime, which is a plausible hypothesis taking into account that the electrical current also present higher values (see {Figs.~\ref{WVforms}}(a) and {\ref{piVSfRate}}(a)) in this configuration.

In summary, when the APPJ impinges on the copper target both $T_{rot}$ and $T_{vib}$ are always smaller for the mesh configuration they tend to not differ significantly when the configuration is changed and when the target is the glass plate. Furthermore, the $T_{rot}$ and $T_{vib}$ values obtained in this work are in agreement with others reported in the literature for APPJs that employ Ar as the working gas \cite{Bazavan_confirmation_2017,Li_filamentary_2019,Yu_influence_2021}.

\subsection{Effects on the surface of polymer}\label{section:PolymerMod}

The effects of plasma treatment on the surfaces of PP samples using both wire and mesh configurations were evaluated through analysis of the surface morphology, measurements of the water contact angle (WCA) and X-ray photoelectron spectroscopy (XPS) analysis. In all cases, the treated and non-treated samples were compared. In samples that were submitted to APPJ treatment the plasma jet was directed at normal incidence to a position close to their center. In all cases studied in this section, the working gas employed in the plasma treatments was Ar at a flow rate of 2.0 l/min, and the distance between the plasma outlet and the sample was 6 mm in both configurations.

\subsubsection{WCA analysis}\label{section:WCAstudy}
\begin{figure}[htb]
\centering
\includegraphics[width=10 cm]{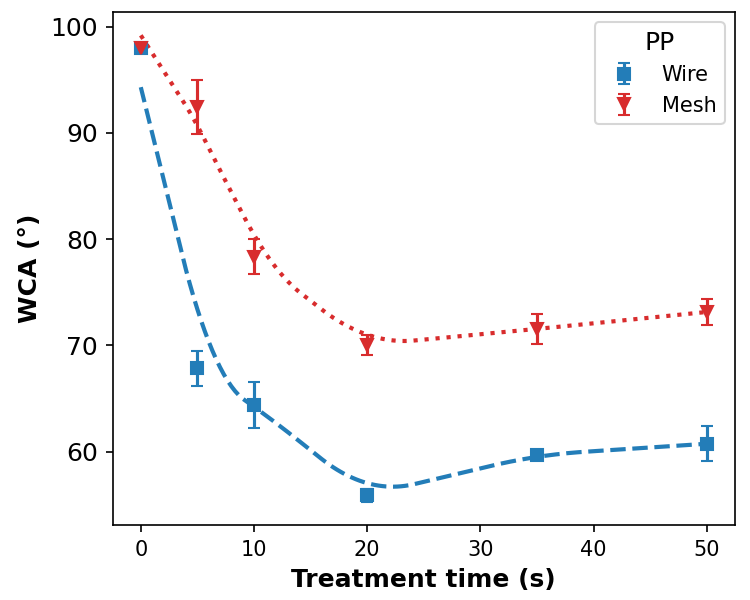}
\caption{WCA measurements as a function of the APPJ treatment time. \label{wcaVStime}}
\end{figure}

The effect of the APPJ exposure on the WCA of PP samples was analyzed as a function of treatment time and the results are shown in Fig.~\ref{wcaVStime}. From Fig.~\ref{wcaVStime}, it is clear that the plasma jet produced with the wire configuration is able to reduce the WCA values faster than the plasma jet obtained with the mesh, and also makes WCA values smaller in the treatment period under analysis.

The behavior of the WCA values as a function of the treatment time is almost the same in both wire and mesh configurations, presenting an abrupt decrease for treatment times up to 20 seconds followed by a slightly increasing plateau. 
Those increments in the WCA values for exposure times higher than 20 s shown in Fig.~\ref{wcaVStime} suggests that an etching of the functional groups created on the surfaces of the samples during APPJ treatment may be occurring.

The measured values of the discharge power when the plasma jets were applied to PP samples were $P_{dis}$ = (3.1 $\pm$ 0.2) W for the wire configuration and $P_{dis}$ = (2.7 $\pm$ 0.3) W for the mesh one. The rotational and vibrational temperature values measured when the sample was a dielectric material were $T_{rot} \approx$ 600 K and $T_{vib} \approx$ 2300 K for wire and $T_{rot} \approx$ 500 K and $T_{vib} \approx$ 1900 K for mesh. Taking into account that all these parameters have higher values when the wire configuration is employed on the material treatments, one of the explanations for the higher reduction in the WCA values is related to a synergistic effect among such parameters. However, the shorter discharge duration observed when the mesh configuration is employed must also be considered, since this effectively reduces the APPJ exposure time of the material under treatment. From Fig.~\ref{wcaVStime} it can be seen that the WCA values for treatment times equal or greater than 10 s obtained with the mesh mounting is nearly 20$\%$ higher than those obtained for the wire case. On the other hand, the discharge duration when using the wire configuration is $\sim$35$\%$ higher than the discharge duration of the mesh one, which correlates with the lower WCA values obtained in this case.

It is worth to mention at this point that basic temperature measurements were performed on a copper target in order to check if the heat transfer from plasma to target would be able to warm significantly the material under plasma exposure. In the set of measurements we executed for that purpose, the highest temperature variation observed on the copper plate was close to five Celsius degrees (from 24 $\degree$C to 28.8 $\degree$C) after impinging the target with the APPJ for 10 minutes. This measurement was carried out using the wire configuration with an Ar flow rate of 1.0 l/min (the $T_{rot}$ value was $\sim$1500 K in this case). We also performed the same measurement employing a flow rate of 2.0 l/min (the $T_{rot}$ value was $\sim$1200 K in this condition). In this case, the temperature variation on the target surface was lower than three Celsius degrees. Moreover, no significant warming is noticed at the exact point where the plasma touches the copper target when it is touched with the bare finger.

\subsubsection{XPS analysis of APPJ treated PP samples}\label{section:XPSstudy}

The surface chemical composition of the PP samples was investigated by XPS analysis. The obtained results are shown in tables~\ref{xpsFractions} to \ref{xpsC1s}. Table~\ref{xpsFractions} show of the element fractions detected on the sample surfaces before and after APPJ treatment using the wire and mesh configurations. In both cases were observed reductions in the carbon (C) content on the surfaces which are accompanied by increments in the oxygen (O) content. Small amounts of nitrogen (N) were also detected in the samples after APPJ treatment. When the mesh configuration was employed, traces of fluorine (F) were also detected on the PP surface, probably coming from the PTFE tube due to the plasma-material interaction.

\begin{table}[htb]
\caption{Averaged elemental composition, in percentage, of PP samples before and after plasma treatment.}\label{xpsFractions}
\centering
\begin{tabular}{lccc}
\hline
\hline
{} & \textbf{C ($\%$)} & \textbf{O ($\%$)} & \textbf{N ($\%$)}\\
\hline
Untreated & 98.8 & $\sim$0.2 & $\sim$0.9\\
Wire & 80.7 $\pm$ 0.9 & 17.8 $\pm$ 0.9 & 1.4 $\pm$ 0.1\\
Mesh & 86.7 $\pm$ 0.8 & 12.3 $\pm$ 0.8 & 0.72 $\pm$ 0.06\\
\hline
\hline
\end{tabular}
\end{table}

In Table.~\ref{xpsRatios} are shown the element ratios N/C and O/C, in percentage. The O/C ratios make more evident the reduction in the C fraction and the increment in the O fraction shown in Table.~\ref{xpsFractions}.
The N/C ratios shown in Table.~\ref{xpsRatios} do not change in the same proportion as in the O/C ratios. However, they show clearly that the insertion of N into the PP surfaces is higher when the wire configuration is used.

\begin{table}[htb]
\caption{Comparison between average element ratios, in percentage, resulting from APPJ treatments with wire and mesh configurations.}\label{xpsRatios}
\centering
\begin{tabular}{lcc}
\hline
\hline
{} & \textbf{O/C ($\%$)} & \textbf{N/C ($\%$)}\\
\hline
Untreated & $\sim$0.91 & $\sim$0.21\\
Wire & 22.4 $\pm$ 1.3 & 1.8 $\pm$ 0.2\\
Mesh & 14.3 $\pm$ 1.0 & 0.84 $\pm$ 0.08\\
\hline
\hline
\end{tabular}
\end{table}

Finally, Table.~\ref{xpsC1s} shows the bindings in C 1s peak obtained after APPJ treatment for the components C-C/C-H, C-OH/R, C=O and COOH/R, whose binding energies are 285.0 eV, 286.7 eV, 287.8 eV and 289.2 ± 0.2 eV, respectively. The APPJ treatments performed with both wire and mesh configurations present reductions in the C-C/C-H components and increments in the C-OH/R, C=O and COOH/R ones, as a consequence of the reduction in the C fraction and increment of O fraction, respectively. The XPS results shown in tables ~\ref{xpsFractions} to \ref{xpsC1s} proves the insertion of O-containing functional groups (hydroxyls, carbonyls, carboxyls) on the surface of the PP samples as the O-fraction is increased compared to the non treated PP-material.

Concerning the differences in the amount of functional groups detected on the PP surfaces after APPJ treatment with the different configurations, we can see that the wire mounting provides a higher reduction of the C-C/C-H components and a more efficient insertion of O-containing ones compared to the employment of the mesh mounting.

\begin{table}[htb]
\caption{Comparison of average bindings in C 1s peaks, in percentage, resulting from APPJ treatments with wire and mesh configurations.}\label{xpsC1s}
\centering
\begin{tabular}{lcccc}
\hline
\hline
{} & \textbf{C-C/C-H ($\%$)} & \textbf{C-OH ($\%$)} & \textbf{C=O ($\%$)} & \textbf{COOH ($\%$)}\\
\hline
Untreated & 98.8 & $\sim$1.2 & $\sim$0.0 & $\sim$0.0\\
Wire & 75.6 $\pm$ 1.5 & 9.6 $\pm$ 0.5 & 6.7 $\pm$ 0.4 & 8.0 $\pm$ 0.6\\
Mesh & 85.0 $\pm$ 1.2 & 6.5 $\pm$ 0.4 & 3.8 $\pm$ 0.3 & 4.8 $\pm$ 0.4\\
\hline
\hline
\end{tabular}
\end{table}

By analyzing the set of results obtained through XPS measurements, we can safely state that the presence of O-containing functional groups, together with the reduction of C-C/C-H ones, on the PP surfaces are the main responsible for the WCA reduction presented in Fig.~\ref{wcaVStime}.

\subsubsection{Analysis of the surface morphology}\label{section:Morphology}

\begin{figure}[htb]
\centering
\begin{tabular}{ccccc}
\begin{subfigure}{0.29\textwidth}
\centering
\includegraphics[width=\textwidth]{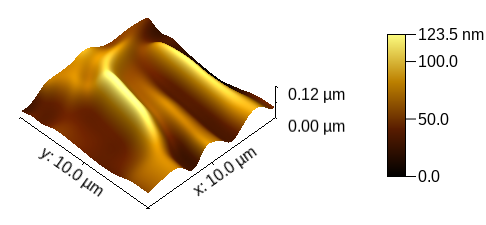}
\subcaption{Untreated sample - $R_q$ = 24 $\pm$ 9 nm}
\end{subfigure}
&
&
\begin{subfigure}{0.29\textwidth}
\centering
\includegraphics[width=\textwidth]{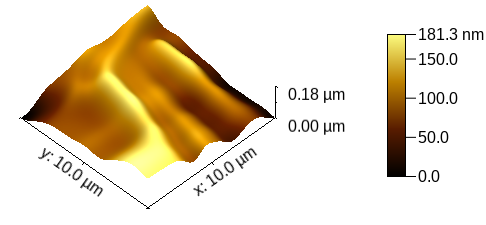}
\subcaption{Wire mounting treated - $R_q$ = 35 $\pm$ 8 nm}
\end{subfigure}
&
&
\begin{subfigure}{0.29\textwidth}
\centering
\includegraphics[width=\textwidth]{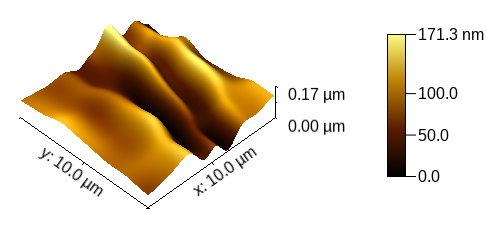}
\subcaption{Mesh mounting treated - $R_q$ = 32 $\pm$ 3 nm}
\end{subfigure}
\end{tabular}
\caption{Surface morphologies of PP samples before and after plasma treatment.\label{surfMorph}}
\end{figure}

In Fig.~\ref{surfMorph} are shown microphotographs of the morphology of PP samples measured using an optical profiler together with the corresponding RMS roughness ($R_q$) values obtained before and after plasma treatment. The profilometry performed on the samples shown in {Fig.~\ref{surfMorph}} was carried out in an area of 10 $\upmu$m x 10 $\upmu$m. In addition, profilometry measurements were also performed scanning an area of 200 $\upmu$m x 200 $\upmu$m. The $R_q$ values obtained for this larger area were 27 $\pm$ 7 nm for the non treated sample, 29 $\pm$ 6 nm after treatment using the wire configuration and 33 $\pm$ 10 nm for the mesh one.
By observing the surface profiles and the corresponding $R_q$ value obtained for each sample, it can be seen that the surface morphology is not much affected by APPJ treatment. Concerning the employment of the different configurations on the modification of the surfaces, it can be seen that both wire and mesh configurations tend to generate an increment in the surface roughness.

From Fig.~\ref{surfMorph} and from the $R_q$ values obtained for the PP-material it can be inferred that the APPJ treatments do not cause a substantial physical damage on the samples because the surface morphologies do not present drastic changes when comparing the treated and non-treated materials. This is a point that differs from what is found in the literature, since most works have been reported significant changes in the surfaces of the PP material. Usually, APPJ treatments introduce small scale roughness on the material surfaces. \cite{kostov_surface_2014, resnik_extracellular_2020, turicek_investigation_2021}. A possible explanation for this difference in relation to what is usually reported in the literature may be related to the higher density of the materials used in this study. This makes the surface hardness be higher, which difficult morphological modifications by APPJ treatment.

\section{Conclusions}

In this work we report on a different design concept to generate APPJs using long and flexible tubes in an assembly that uses a metallic mesh between two plastic tubes in a concentric arrangement. The plasma jet parameters obtained with this configuration and the application on treatment of polymeric materials were compared to those from a device mounting that employs a long tube with a metal wire inside it. Both arrangements work using the jet transfer technique.
Taking into account the $P_{dis}$ values obtained in the wire or the mesh configurations, it is safe to say that the wire one is better for applications which require higher $P_{dis}$ values. However, considering the $i_{RMS}$ values measured in the different configurations, the mesh one would be the most suitable choice for application that require lower electrical current values, like the \textit{in vivo} ones.

Concerning the thermal parameters of the APPJs obtained with the wire and mesh configurations, the $T_{rot}$ values measured with are significantly lower when the mesh mounting is employed and the target is copper. Since $T_{rot}$ has a close relationship with $T_{gas}$, the use of the mesh mounting seems to be more appropriated for applications in sensitive materials that present significant conductivity.

Regarding the production of reactive species, the relative production of excited OH tends to be higher when the mesh configuration is employed. The relative production of excited NO does not differ significantly when the configuration is changed and the APPJ impinges a metallic target. On the other hand, when the target is a dielectric material, the production of NO species present a tendency to be higher when the wire mounting is employed. Thus, taking into account the differences in the relative production of OH and NO, the choice between wire and mesh configurations must take into account the need for larger amounts of one species or low rms current and power.

The results on the polymer treatment using both configurations have shown that the set of parameters obtained for the plasma jet in the wire configuration lead to better results in terms of WCA reduction and insertion of O-containing functional groups on the surface of the PP material. Also, the WCA and XPS measurements revealed that the APPJ treatment insert O-containing functional groups on the polymer surfaces, which is the main reason for the WCA reduction.

An interesting point regarding the employment of the mesh design is that the mesh itself could be replaced by a metallic film covering the outer wall of the inner tube. This possibility would allow the use of inner tubes with very small diameters, which in turn would provide a very narrow plasma jet that would fit well in dental treatments for example, which require the application of APPJs in very specific and small areas. Of course, this configuration can also be employed in APPJ devices that do not use the jet transfer technique.

In a future work, we intend to investigate the longitudinal distribution of the APPJ treatment on PP and also on other polymer surfaces. Moreover, we intend to perform a detailed study of how a plasma jet propagates in long tubes employing the mesh configuration.

\section*{Acknowledgements}
The authors thank the Leib­niz In­sti­tu­te for Plas­ma Sci­ence and Tech­no­lo­gy (INP) for the access to the XPS facility, the Center for Semiconducting Components and Nanotechnologies (CCSNano) for providing the surface morphology measurements and Bruno Silva Leal for technical support in this work. This research was supported by FAPESP (grants $\#$2019/05856-7 and $\#$2020/09481-5).

\section*{Data Availability Statement}
Data are contained within the article. Raw data are available from the authors under reasonable request.

\bibliographystyle{unsrt}
\bibliography{Transfered_and_long_tube_plasma_jets}

\end{document}